\documentclass[
    ,final            % use final for the camera ready runs
  ]
  {aipproc}

\layoutstyle{6x9}

\begin{document}

\title{Gamma-ray Bursts as Dark Energy Probes\footnote{\uppercase{T}alk presented 
by O.B. at {T}he {D}ark {S}ide of the {U}niverse {I}nternational 
{W}orkshop, {M}adrid, Spain, 20--24 {J}une 2006.}
}



\author{O. Bertolami}{
}

\author{P.T. Silva}{
  address={Instituto Superior T\'ecnico, Departamento de F\'\i sica\\
Av. Rovisco Pais, 1049-001\\ 
Lisboa, Portugal\\ 
E-mail: orfeu@cosmos.ist.utl.pt; paptms@ist.utl.pt}
}

\begin{abstract}
We discuss the prospects of using Gamma Ray Bursts (GRBs) as
high-redshift distance estimators, and consider their use in the study of two
dark energy models, the Generalized Chaplygin Gas (GCG), a model for the unification 
of dark energy and dark matter, and the XCDM model, a model where a generic dark
energy fluid like component is described by the equation of state, 
$p= \omega \rho$. We find that this test yields rather disappointing
results for the GCG model, being mainly sensitive to the total amount
of matter present in the Universe in the case of the XCDM model. 
We also find that, within the framework of the XCDM model, a large
sample of GRBs ($\geq 200$) may turn out to be quite useful to improve the
forthcoming type Ia supernovae data. 

\end{abstract}

\maketitle

\section{Introduction}

Recently, there has been a great deal  of activity on attempts of
using GRBs as cosmological probes~\cite{bertolami06}. In the original 
proposal~\cite{schaefer03}, it has been suggested that
the magnitude versus redshift plot could be extended
to redshifts up to $z\simeq 4.5$, via correlations found
between the isotropic equivalent luminosity, $L_{iso}$, and two GRB
observables, namely the time lag ($\tau_{lag}$)~\cite{norris00} and
variability ($V$)~\cite{reichart01}. The isotropic equivalent luminosity
is the inferred luminosity (energy emitted per unit of time) of a GRB if all
its energy is radiated  isotropically,
the time lag measures the time offset between high- and low-energy arriving
GRB photons, while the variability is a measure of the
complexity of the GRB light curve. Using these correlations, one can infer
two estimates of the absolute isotropic equivalent luminosity, which are 
combined through a weighted average. Knowing the absolute isotropic equivalent
luminosity together with the observed fluence yields an estimate of the luminosity
distance to the GRB.

Unfortunately, these correlations are
affected by a large statistical (or intrinsic) scatter. This statistical
spread affects not only the cosmological precision via its direct statistical
contribution to the distance modulus uncertainty, $\sigma_\mu$, but also through 
the calibration uncertainty given that the suitable GRB sample with known redshift is
rather small. 
In what follows we show that a relatively small sample of GRBs with low redshifts is
sufficient to greatly reduce the systematic uncertainty thanks to a more robust
and precise calibration.

More recently, a new correlation has been suggested~\cite{ghirlanda04a}, which is
subjected to a much smaller statistical scatter. 
The so-called Ghirlanda relation is a correlation between the
peak energy of the gamma-ray spectrum, $E_{peak}$ (in the $\nu - \nu F_{\nu}$
plot), and the collimation-corrected energy emitted in gamma-rays,
$E_\gamma$. This collimation-corrected energy is a measure of the energy
released by a GRB taking into account that it is beamed into a 
narrow jet. Unlike the $L_{iso}-\tau$ and $L_{iso}-V$ relations, the
Ghirlanda relation is not affected by large statistical
uncertainties, however, it depends on poorly constrained quantities
related to the
properties of the medium around the burst. Indeed, to infer $E_\gamma$ one
must estimate the  angular opening of the jet, which can be performed
assuming a density profile for the medium around the burst
(or circum-burst medium for short), where a fraction $\eta_\gamma$ of the
fireball kinetic energy is emitted in the prompt gamma-ray phase, and where
one has measured the jet break time, $t_{jet}$ \cite{sari99}. Assuming 
that the circum-burst medium has a constant density, the simplest possible
assumption, requires one additional parameter. This constant density
has been measured for a few bursts~\cite{ghirlanda04a}, and it
exhibits a wide variation from burst to burst.

Another difficulty involving GRBs is that they tend to occur at rather large distances,
which makes it impossible to calibrate any relationship between the relevant variables 
in a way that is independent from the cosmological model. The method that is 
usually employed consist in fitting both, the cosmological \emph{and} the 
calibration parameters, and then use statistical techniques to remove the
undesired parameters. In here, we follow a different procedure
\cite{bertolami06,takahashi}. 
We consider that the luminosity distance for $z<1.5$ was previously measured
using type Ia supernovae, and divide the GRBs sample in two sets; the low redshift
sample, with  $z<1.5$, and the high redshift one, with $z>1.5$.
Since the luminosity distance of GRBs in the range $z<1.5$ is
already known, one can 
calibrate the luminosity estimators independently of the cosmological
parameters and use the high redshift sample as a probe to dark energy and dark matter
models. This method also allows us to verify whether the larger redshift range
probed by GRBs can compensate for the larger uncertainty associated with the
distance estimates thus obtained.

We have analyzed the use of these correlations in order to study of GCG, a
model that unifies the dark energy and dark matter in a single fluid
\cite{bento02} through the equation of state 
$ p_{ch} = - A / \rho_{ch}^\alpha$, where $A$ and $\alpha$ are positive
constants. The case $\alpha=1$ describes the 
the Chaplygin gas, that arises in different theoretical scenarios.
If the curvature is fixed, there are only two free variables,
$A$ and $\alpha$, although it is more convenient to use the quantity
$A_s \equiv A/ \rho_{ch,0}^{1+\alpha}$ instead of $A$. Thus, we consider
two free parameters, $\alpha$ and $A_s$. A great deal of effort has been 
recently devoted to  constrain the GCG model parameters~\cite{bertolami05},
which include, for instance, supernovae~\cite{bertolami04,bento05b},
cosmic microwave background radiation~\cite{bento03a,bento03b,bento03c},
gravitational lensing~\cite{silva03} and cosmic topology~\cite{bento05a}.

In addition to the GCG model, we also study the more conventional
flat XCDM model. Likewise the GCG model, the
XCDM model is also described by two free parameters,
the  dark energy equation of state, $\omega \equiv p / \rho$, and the fraction
of non-relativistic matter, $\Omega_m$. Testing these models is particularly
relevant since they are degenerate for redshifts 
$z<1$~\cite{bertolami04,bento05b}.

\section{Variability and Time Lag as Luminosity Estimators}

Let us describe here the approach based on the correlation between the
isotropic luminosity, and the variability and time lag.
The $L_{iso}-\tau$ and $L_{iso}-V$ correlations are written as
\begin{equation}
L_{iso}=B_\tau \tau_{lag}^{\beta_\tau}~, ~~~~~~
L_{iso}=B_v V^{\beta_v}~.
\end{equation}
The parameters $B_{\tau/v}$ and $\beta_{\tau/v}$ are found through
fitting of these relationships to the data points, that is, via
calibration of the luminosity estimators. Given that the GRB sample
with measured redshifts is rather small, at present the calibration is rather poor.
To test what improvements one expects to achieve in the future, 
we assess the gain in  calibrating these relationships with larger samples.
Three mock, yet realistic, samples were generated, using the method
detailed in Ref.~\cite{bertolami06}; these mock data sets were used
to calibrate the luminosity estimators.

We find that a calibration performed with $40$ GRBs greatly improves the
previous results, decreasing $\sigma_\mu$ by close to half, yielding
$\sigma_\mu=0.68$. However, by increasing the 
calibration sample to $100$ GRBs the resulting improvement is just
marginal, suggesting
that very large calibration samples are not required. It is worth noting
that a sample of about $40$ GRBs may be available in the near future thanks
to the {\it Swift} satellite. We also find that despite the large statistical
scatter, due to improvements in calibration, the uncertainty for this estimator
becomes quite close to that obtained with the Ghirlanda relation, 
$\sigma_\mu = 0.5$ (c.f. below).
However, it is also evident that, due to this large statistical uncertainty,
one cannot expect to significantly reduce the observational uncertainty of the
variability and time lag method beyond about $\sigma_\mu \approx 0.6$. 
In a sense, one arrives at a minimum possible uncertainty plateau, beyond
which any further improvement seems impossible.

\begin{figure}[t]
  \includegraphics[width=\textwidth]{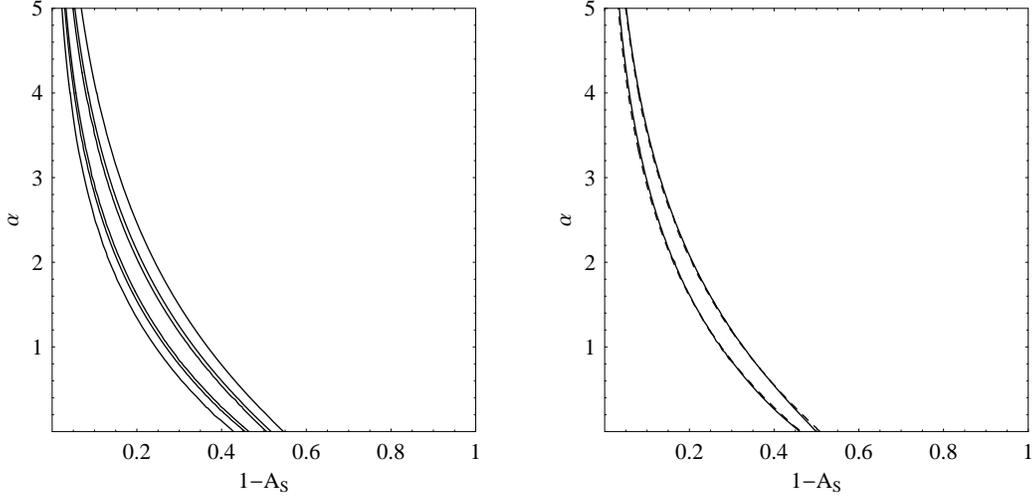}
  \caption{Encountered confidence regions for the GCG model. The figure on the
  left shows the effect of increasing the number of GRBs in the sample.
  The curves show the 68\% Confidence Level (CL) regions, from the outer
  to the inner curves, corresponding to 150, 500 and 1000 high-redshift
  GRBs. On the right, the solid line shows the 68\% CL regions obtained
  through a sample of 100 low-redshift ($z<1.5$)
  and 400 high-redshift ($z>1.5$) GRBs, while the dashed line show the 68\% CL
  constraints for a sample made up of 500 high-redshift GRBs only. On the left
  figure, the $\tau-L_{iso}$ and $V-L_{iso}$ relations have been used, while on
  the right one the Ghirlanda relation was employed. 
  The degeneracy of the $\alpha$ parameter is quite evident.\label{gcg}}
\end{figure}

Next, we examine how GRBs fare when used to constrain both models
under consideration. The methodology is essentially the same of that
used for type Ia supernovae \cite{silva03}. One starts by defining a 
fiducial cosmological model, described by the parameters
$\mathbf{p}_\mathrm{fid}$,
and then use the $\chi^2$ function, defined as
\begin{equation}
\chi^2(\mathbf{p})= \sum_i^{N_{GRB}} \left[ 5 \log D_L(z_i,\mathbf{p}_\mathrm{fid})
 - 5 \log D_L(z_i,\mathbf{p}) \over \sigma_\mu \right]^2~,
\end{equation}
where $D_L$ is the dimensionless luminosity distance, to build confidence regions
in parameter space. The dimensionless luminosity distance of the GRBs
can be estimated using $L_{iso}-\tau$ and $L_{iso}-V$.
 
Somewhat against our expectation, we found that GRBs are not very suited to study
the GCG model. The $A_s$ parameter can be constrained, however no limit can be
imposed on $\alpha$, as shown in Figure \ref{gcg}. Larger samples of GRBs
decrease the area of the allowed parameter space, however there is no
significant improvement on the constraints imposed on either parameter.
We also find that using some low-redshift GRBs or the Ghirlanda relation
does not alter these conclusions significantly.

\begin{figure}[t]
  \includegraphics[width=\textwidth]{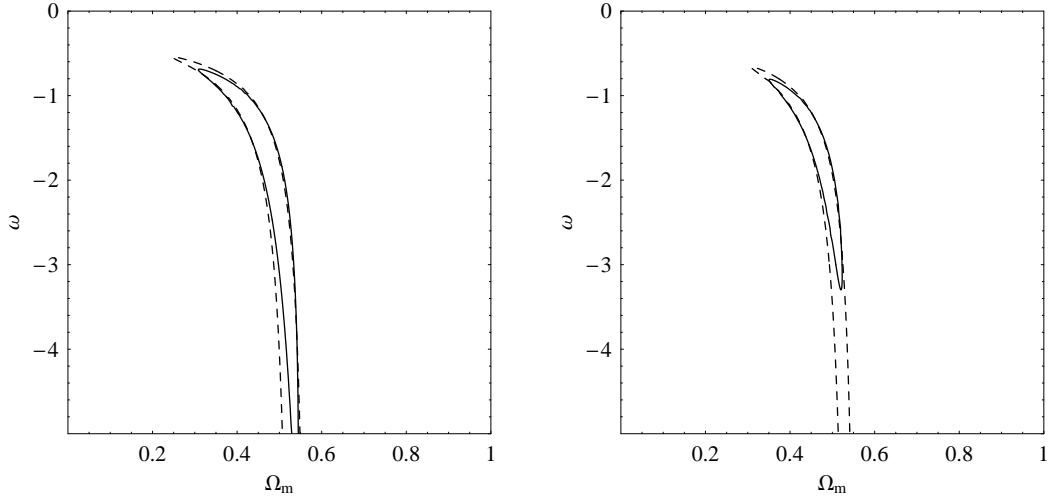}
  \caption{Encountered confidence regions for the XCDM model. The solid lines show
  the 68\% CL regions obtained through a sample of 100 low-redshift ($z<1.5$)
  and 400 high-redshift ($z>1.5$) GRBs, while the dashed lines show the 68\% CL
  constraints for a sample made up of 500 high-redshift GRBs only. On the left
  figure, the $\tau-L_{iso}$ and $V-L_{iso}$ relations have been used, while on
  the right one the Ghirlanda relation was employed. \label{xcdm}}
\end{figure}

The results for the XCDM model are, however, more promising.
We find that GRBs are sensitive essentially to $\Omega_m$,
and very weakly sensitive to $\omega$. The reason for this is the
redshift range probed by GRBs. We have verified
that when using a sample that includes 100 GRBs with $z<1.5$, the constraints
on the XCDM model are substantially better, as depicted in Fig. \ref{xcdm}. 
It should be pointed out that this redshift dependence is not found for
the GCG fiducial model. These results were found using the minimal
$\sigma_\mu=0.66$. This uncertainty is essentially due to the statistical
component, and hence it cannot be reduced by better calibration or data,
only through larger GRB samples.

\section{The Ghirlanda Relation.}

We discuss here the use of the Ghirlanda relation, which is known to be
intrinsically more precise. A drawback of this relation is its dependence
on more parameters, and on how the circum-burst medium is
modeled~\cite{nava05}. The calibration testing procedure was not repeated,
as the main sources of uncertainty in the Ghirlanda relation are the poorly
constrained values of the peak energy,
jet break time and circum-burst density~\cite{silva03}.

As before, we find that the characteristic feature of GRBs of having rather high
redshifts, makes them somewhat unsuitable to study dark energy models, even the
GCG one. Despite the increased precision, the allowed parameter range for the
GCG model is not greatly improved when one uses the Ghirlanda relation.
As for the XCDM model, one finds that results are better if one uses
the Ghirlanda relations, but not significantly in what concerns the 
dark energy component. However, this independence on the the nature and amount
of the dark energy component means that GRBs can provide a estimate of $\Omega_m$
alone, something which is not possible when using type Ia supernovae.

Thus, while an improvement in calibration should not greatly alter the
above conclusions, it should be noted that data quality and statistics will
greatly improve in the future thanks to {\it Swift} and {\it HETE 2}
experiments. Thus, as one expects  significant improvements on the
determinations of the peak energy, jet break time and circum-burst density,
it is reasonable to assume that the distance modulus uncertainty for
the Ghirlanda relation will decrease.

\section{Final Remarks.}

\begin{figure}[t]
  \includegraphics[width=\textwidth]{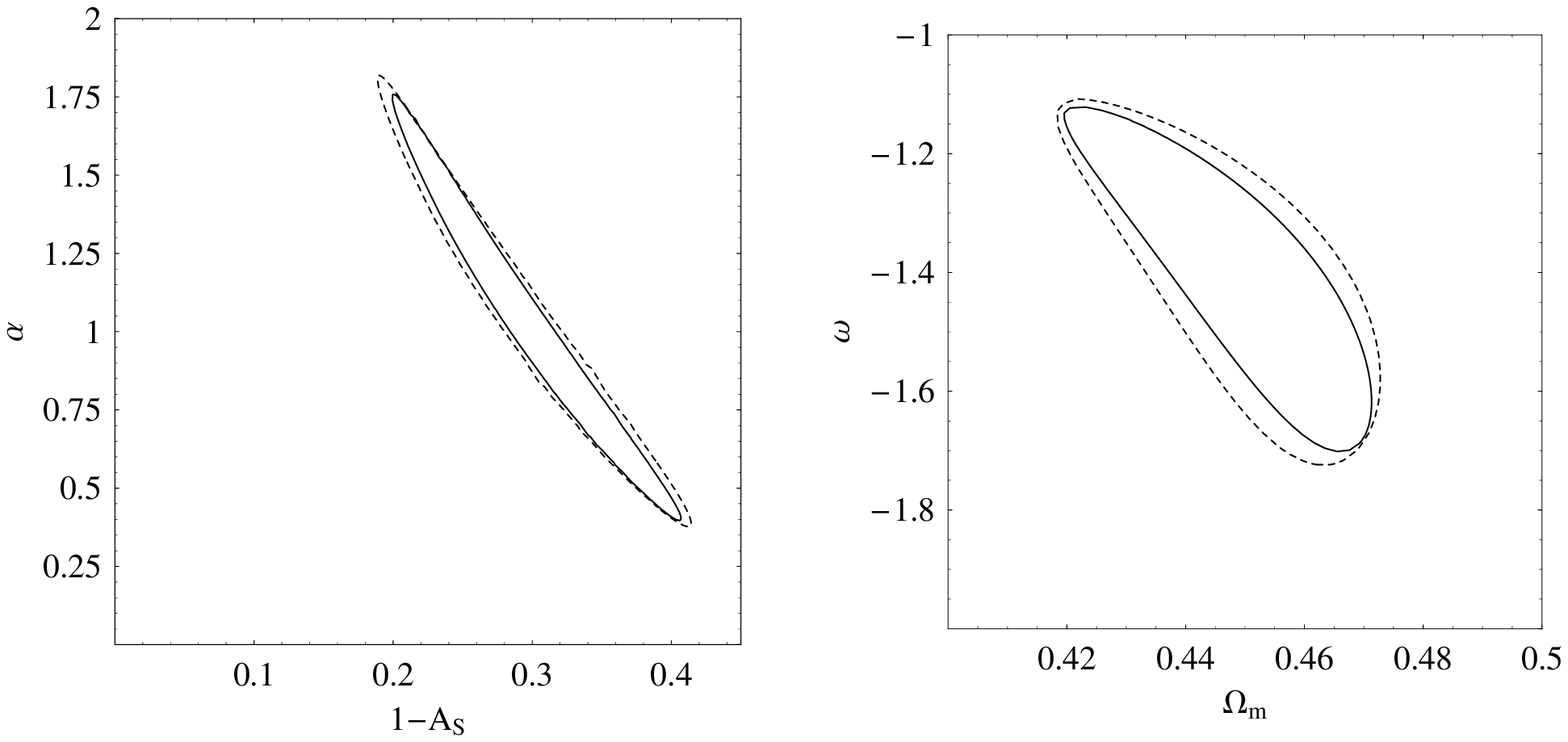}
  \caption{Joint constraints from SNAP plus 500 high-redshift GRBs for the
  CGC(left) and XCDM (right) model. The dashed line corresponds only to SNAP
  constrains, while the solid region corresponds the to the SNAP+GRBs ones.
  All curves correspond to 68\% CL. Notice that an improvement,
  although marginal, is obtained.  \label{SNGRB}}
\end{figure}

The main conclusion of our study is that although GRBs are poor dark energy
probes, for $z>1.5$, their luminosity distance is quite sensitive to the
dominating energy density component. For the XCDM model, this is dark matter,
and we find
that the amount of dark matter can be remarkably constrained. For the GCG
model, on the other hand, it turns out that what arises is a combination of
the $A_s$ and $\alpha$ parameters, and the data cannot lift the degeneracy
on $\alpha$. Actually, if $z \gg 0$, the Hubble function for the GCG becomes
\begin{equation}
H_{ch}(z \gg 0) =  \Omega_{ch}(1-A_s)^{ 1 /( 1 + \alpha ) }(1+z)^3~,
\end{equation}
and one can easily observe that the allowed parameter region predicted
from GRBs does follow the line $\Omega_{ch}(1-A_s)^{ 1 /( 1 + \alpha ) }
\approx 0.5$. This feature is also encountered in various phenomenological
studies of the GCG, the only exception being on data from large scale structure
formation~\cite{bento04}. Also, the transition into an accelerated expansion
phase in a GCG universe is faster, and at a lower redshift, than for the
XCDM model. This explains why using some $z<1.5$ GRBs improves the
results for the latter model, while it does not have any impact on the former.

The sensitivity on the non-relativistic matter density in XCDM models means
that GRBs may have a complementary role to play with respect to type Ia
supernovae. To test the impact of such a joint use, we built joint confidence
regions for GRBs and type Ia supernovae constraints that may be imposed by the
\textit{SNAP} 
satellite. It was found that GRBs may play an important role in the near
future if a large GRB data set is built before the promised scientific bounty of 
\textit{SNAP} becomes available (see Figure \ref{SNGRB}). However, GRBs should only
marginally improve the constraints imposed by \textit{SNAP}, unless the uncertainty
is reduced by  at least a factor of two by then.

It must be realized that these results did not take into account several other
systematic sources of error, namely selection and gravitational
lensing effects. While for a tentative study, such as the one considered
in this contribution, these potential sources of uncertainty may be
neglected, a more careful assessment must be performed 
if one aims to impose robust constraints on cosmological models. Furthermore,
it is relevant to point out that the used correlations are purely phenomenological
and lack, so far, a theoretical explanation.

It is interesting that a new correlation has been recently
proposed~\cite{liang05} which does not require any assumptions with 
regards to the circum-burst medium or the gamma-ray production efficiency.
We are currently in the process of assessing the potential of
such a relationship and testing whether 
marginalization methods, such as those used for type Ia supernovae, can be
advantageously employed~\cite{bertolami07}.

\end{document}